\documentclass[journal,comsoc]{IEEEtran}
\usepackage[T1]{fontenc}

\ifCLASSINFOpdf
\else
\fi
\usepackage[cmex10]{amsmath}
\usepackage[cmintegrals]{newtxmath}

\ifCLASSINFOpdf
   \usepackage[pdftex]{graphicx}
\else
   \usepackage[dvips]{graphicx}
\fi

\usepackage[caption=false,font=footnotesize]{subfig}

\usepackage{listings}
\usepackage{color}
\definecolor{dkgreen}{rgb}{0,0.6,0}
\definecolor{gray}{rgb}{0.5,0.5,0.5}
\definecolor{mauve}{rgb}{0.58,0,0.82}
\usepackage{bm}
\usepackage{algorithm}
\usepackage{algorithmicx}
\usepackage{algpseudocode}

\usepackage{booktabs}
\usepackage[caption=false]{subfig}
\usepackage{multirow}
\usepackage{authblk}
\newcommand{\NA}{---}

\begin{document}
%
\title{Android HIV: A Study of Repackaging Malware for Evading Machine-Learning Detection}
%
%
%

\author{Xiao Chen,
        Chaoran Li, Derui Wang, Sheng Wen, Jun Zhang, Surya Nepal, Yang Xiang, and Kui Ren
\thanks{Manuscript received December 12, 2018; revised May 29, 2019; accepted July 15, 2019. Date of publication July 31, 2019; date of current version October 8, 2019. The associate editor coordinating the review of this manuscript and approving it for publication was Prof. Loukas Lazos. (Corresponding author: Xiao Chen.)}
\thanks{X. Chen, C. Li, D. Wang, S. Wen, J. Zhang, and X. Yang are with the Faculty of Science, Engineering and Technology, Swinburne University of Technology, Hawthorn,
VIC 3122, Australia. E-mail: xiaochen@swin.edu.au.}
\thanks{S. Nepal is with Data61, CSIRO, Australia.}
\thanks{K. Ren is with Department of Computer Science and Engineering, University at Buffalo, State University of New York, Buffalo, NY 14260, USA.}
\thanks{Digital Object Identifier 10.1109/TIFS.2019.2932228}
}

%
%

\markboth{IEEE TRANSACTIONS ON INFORMATION FORENSICS AND SECURITY,~Vol.~15, 2020}%
{Chen \MakeLowercase{\textit{et al.}}: Android HIV: A Study of Repackaging Malware forEvading Machine-Learning Detection}
%



\maketitle

\begin{abstract}

Machine learning based solutions have been successfully employed for automatic detection of malware on Android. However, machine learning models lack robustness to adversarial examples, which are crafted by adding carefully chosen perturbations to the normal inputs. So far, the adversarial examples can only deceive detectors that rely on syntactic features (\textit{e.g.}, requested permissions, API calls, \textit{etc.}), and the perturbations can only be implemented by simply modifying application's manifest. While recent Android malware detectors rely more on semantic features from Dalvik bytecode rather than manifest, existing attacking/defending methods are no longer effective. 

In this paper, we introduce a new attacking method that generates adversarial examples of Android malware and evades being detected by the current models. To this end, we propose a method of applying optimal perturbations onto Android APK that can successfully deceive the machine learning detectors. We develop an automated tool to generate the adversarial examples without human intervention. In contrast to existing works, the adversarial examples crafted by our method can also deceive recent machine learning based detectors that rely on semantic features such as control-flow-graph. The perturbations can also be implemented directly onto APK's Dalvik bytecode rather than Android manifest to evade from recent detectors. We demonstrate our attack on two state-of-the-art Android malware detection schemes, MaMaDroid and Drebin. Our results show that the malware detection rates decreased from $96\%$ to $0\%$ in MaMaDroid, and from $97\%$ to $0\%$ in Drebin, with just a small number of codes to be inserted into the APK.

\end{abstract}

\begin{IEEEkeywords}
android malware detection, adversarial machine learning.
\end{IEEEkeywords}

%
\IEEEpeerreviewmaketitle

\section{Introduction}
\label{intro}
%
%
%
%
\IEEEPARstart{W}{ith} the growth of mobile applications and their users, security has increasingly become a great concern for various stakeholders. According to  McAfee's report \cite{M2017}, the number of mobile malware samples has increased to 22 millions in third quarter of 2017. 
Symantec further reported that in Android platform, one in every five mobile applications is actually malware \cite{PW2015}. Hence, it is not surprising that the demand for automated tools for detecting and analysing mobile malware has also risen. 
Most of the researchers and practitioners in this area target Android platform, which dominants the mobile OS market.
To date, there has been a growing body of research in malware detection for Android. Among all the proposed methods \cite{PF2015}, machine learning based solutions have been increasingly adopted by anti-malware companies \cite{NA2017} due to their anti-obfuscation nature and their capability of detecting malware variants as well as zero-day samples. 
Despite the benefits of machine learning based detectors, it has been revealed that such detectors are vulnerable to adversarial examples \cite{papernot2016limitations,Carlini2017}. Such adversarial examples are crafted by adding carefully designed perturbations to the legitimate inputs that force machine learning models to output false predictions \cite{Goodfellow2014Explaining,papernot2016limitations,CS2013}.

Analogously,  adversarial examples for machine learning based detection are very much like the HIV which progressively disables human beings' immune system. 
We chose malware detection over Android platform to assess the feasibility of using adversarial examples as a core security problem. In contrast to the same issue in other areas such as image classification, the span of acceptable perturbations is greatly reduced: an image is represented by pixel values in the feature space and the adversary can modify the feature vector arbitrarily, as long as the modified image is visually indistinguishable \cite{XY2017}; however, in the context of crafting adversarial examples for Android malware, a successful case must comply with the following restrictions which are much more challenging than the image classification problem: 1) the perturbation must not jeopardise malware's original functions, and 2) the perturbation to the feature space can be practically implemented in the Android PacKage (APK), meaning that the perturbation can be realised in the program code of an unpacked malware and can also be repacked/rebuilt into an APK.

\begin{figure}[!t]
\centering
\includegraphics[width=0.9\columnwidth]{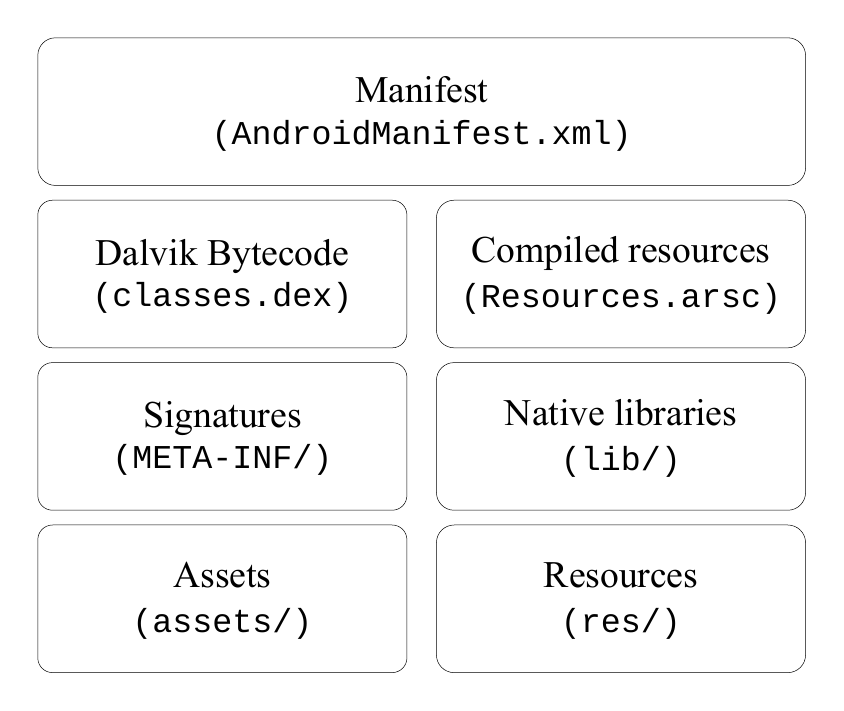}
\caption{File structure of APK. \texttt{AndroidManifest.xml} declares the essential information; \texttt{classes.dex} contains the Dalvik Bytecode; \texttt{resources.arsc} holds the compiled resources in binary format; META-INF, lib, assets, and res folders include the meta data, libraries, assets, and resources of the application, respectively.}
\label{fig_apk}
\end{figure}

So far, there are already a few attempts on crafting/defending adversarial examples against machine learning based malware detection for Android platform. However, the validity of these works is usually questionable due to their impracticality. For example, Chen et al. \cite{chen2018automated} proposed to inject crafted adversarial examples into the training dataset so as to reduce detection accuracy. This method is impractical because it is not easy for attackers to gain access to the training dataset in most use cases. Grosse et al. \cite{grosse2017adversarial} explored the feasibility of crafting adversarial examples in Android platform, but their malware detecting classifier was limited to Deep Neural Network (DNN) only. They could not guarantee the success of adversarial examples against traditional machine learning detectors such as Random Forest (RF) and Support Vector Machine (SVM). Demontis et al. \cite{demontis2017yes} proposed a theoretically-sound learning algorithm to train linear classifiers with more evenly-distributed feature weights. This allows one to improve system security without significantly affecting computational efficiency. Chen et al. \cite{CL2017} also developed an ensemble learning method against adversarial examples. Yang et al. \cite{YW2017} conducted new malware variants for malware detectors to test and strengthen their detection signatures/models. According to our research, all these ideas \cite{demontis2017yes,CL2017,YW2017} can only be applied to the malware detectors that adopt syntactic features (\textit{e.g.}, permissions requested in the manifest or specific APIs in the source code \cite{wu2012droidmat,arp2014drebin,aung2013permission,peiravian2013machine}). However, almost all recent machine learning based detection methods rely more on the semantic features collected from Dalvik bytecode (\textit{i.e.}, \texttt{classes.dex}). This disables existing methods of crafting/defending adversarial examples in Android platform. Moreover, it is usually simple for existing methods to modify the manifest for the generation of adversarial examples. However, when the features are collected from the bytecode, it becomes very challenging to modify the bytecode without changing the original functionality due to their programmatic complexity. Therefore, existing works are not of much value in providing proactive solutions to the ever-evolving adversarial examples in terms of Android malware variants \cite{CL2017,grosse2017adversarial,chen2018automated,demontis2017yes,YW2017}.

In this paper, we propose and study a highly-effective attack that generates adversarial malware examples in Android platform, which can evade being detected by current machine learning based detectors. 
In the real world, defenders and attackers are always engaged in a never-ending war. To increase the robustness of Android malware detectors against malware variants, we need to be proactive and take potential adversarial scenarios into account while designing malware detectors to achieve creating such a proactive design. The work in this paper envisions an advanced method to craft Android malware adversarial examples. The results can be used for Android malware detectors to identify malware variants with the manipulated features. 
For the convenience of description, we selected two typical Android malware detectors, MaMaDroid \cite{mariconti2016mamadroid} and Drebin \cite{arp2014drebin}.
Each of these two selects semantic or syntactic features to model malware behaviours. 

We summarise the key contributions of this paper from different angles of view as follows:
\begin{itemize}
\item Technically, we propose an innovative method of crafting adversarial examples on recent machine learning based detectors for Android malware (\textit{e.g.}, Drebin and MaMaDroid). They mainly collected features (either syntactic or semantic ones) from Dalvik bytecode to capture behaviors of Android malware. This contribution is distinguishable from the existing works \cite{CL2017,chen2018automated,demontis2017yes,grosse2017adversarial} because can only target/protect the detectors relying on syntactic features.

\item Practically, we designed an automated tool to apply the method to the real-world malware samples. The tool calculates the perturbations, modifies source files, and rebuilds the modified APK. This is a key contribution as the developed tool adds the perturbations directly to APK's \texttt{classes.dex}. This is in contrast to the existing works (\textit{e.g.}, \cite{chen2018automated,grosse2017adversarial}) that simply apply perturbations in \texttt{AndroidManifest.xml}. Although it is easy to implement, they cannot target/protect recent Android malware detectors (\textit{e.g.}, \cite{du2017android,shen2017android}) which do not extract features from Manifest. 

\item We evaluated the proposed manipulation methods of adversarial examples by using the same datasets that Drebin and MaMaDroid ($5879$ malware samples) used \cite{arp2014drebin,viennot2014measurement}. Our results show that, the malware detection rates decreased from $96\%$ to $0\%$ in MaMaDroid, and from $97\%$ to $0\%$ in Drebin, with just a small distortion generated by our adversarial example manipulation method. 

\end{itemize}

The rest of the paper is organised as follows. Section \ref{apk} gives an introduction to Android application packaging which forms the basis for adding perturbations. Section \ref{target_system} presents the details of two typical target Android malware detectors as well as the attack scenarios. Section \ref{sec:attack_on_mamadroid} and \ref{attacking_drebin} show how to craft adversarial examples against MamaDroid and Drebin, respectively, followed by discussions on open issues in Section \ref{discussion}. Related work comes in Section \ref{related_work}, and finally, Section \ref{conclusion} concludes the paper.

\section{Android Application Package} 
\label{apk}

Android applications are packaged and distributed in the form of APK files. The APK file is a \texttt{jar}-like archive that packs the application's dexcode (\texttt{.dex} files), resources, assets, and manifest file. The structure of an APK is shown in Fig.\ref{fig_apk}. In particular, \texttt{AndroidManifest.xml} is designed for the meta-data such as permissions requested, definitions of components like Activities, Services, Broadcast Receivers and Content Providers. \texttt{Classes.dex} is used to store the Dalvik bytecode to be executed on the Android Runtime environment. \texttt{Res} folder contains graphics, string resources,  user interface layouts, \textit{etc}. \texttt{Assets} folder includes non-compiled files and \texttt{META-INF} is to store the signatures and certificates.

The state-of-the-art detectors usually use machine learning based classifiers to categorize applications as either malicious or benign \cite{arp2014drebin,aung2013permission,mariconti2016mamadroid,peiravian2013machine,wu2012droidmat}. Features employed by such classifiers are extracted from the APK archive by performing static analysis on the manifest and dexcode.

Manifest introduces the components of an application as well as its requested permissions. Such information is presented in a binary XML format inside \texttt{AndroidManifest.xml}. 

Contents presented in the manifest are informative, implying the intentions and behaviours of an application. For instance, requesting \texttt{android.permission.SEND\_SMS} and \texttt{android.permission.READ\_CONTACTS} permissions indicate that the application may send text messages to your contacts. Features retrieved from the manifest are usually constructed as a vector of binary values, while each value indicates the presence/absence of a certain element in the manifest.
Dexcode, or Dalvik Bytecode, is the operational code on Android platform. All the Java source codes are compiled and assembled into a single Dalvik Executable (\texttt{classes.dex}). Features extracted from \texttt{classes.dex}, such as Control-Flow-Graph (CFG) and Data-Dependency-Graph (DDG), contains rich semantic information and logical structure of the application. They are usually presented in two forms: 1) the raw sequence of API calls, and 2) the statistic information retrieved from the call graph (\textit{e.g.}, similarity scores between two graphs \cite{du2017android}). Such features are proved to have strong discriminating power for identification of malware. 

To evade being detected by machine learning based detectors, a malware sample has to be manipulated so that the extracted features for the learning systems look benign. Intuitively, the target files to be modified are those from which the features are extracted, \textit{i.e.},  \texttt{AndroidManifest.xml} and/or \texttt{classes.dex}. While both of these files are in binary format and are not readable by human, decompiling tools such as apktool are used to convert them into a readable format. Specifically, the binary XML can be transformed into plain-text XML, and the Dalvik bytecode can be disassembled to smali files, which are more human-friendly as intermediate presentations of bytecode. The processed manifest file and smali files can be edited and reassembled to an APK.

\section{Targeted Systems and Attack Scenarios} 
\label{target_system}

We propose a framework to craft adversarial examples that can evade machine learning based detection. 
Generally, machine learning based malware detection methods leverage two types of features: static and dynamic features. Static features are collected from disassembled APKs. Examples of such features include requested permissions, API call sequences, and control flow graphs. Dynamic features, on the other hand, are collected during the execution of the applications by monitoring their behavior and communication patterns. Since dynamic features are collected by feeding random inputs, it is more challenging to alter dynamic features than static features. Therefore, we target static features in this work, and leave the dynamic case for future work. Specifically, we target two typical solutions which have been widely analysed in this field, \textit{i.e.}, MaMaDroid \cite{mariconti2016mamadroid} and Drebin \cite{arp2014drebin}. The semantic features that MaMaDroid uses are extracted from dexcode, and the syntactic string values which are adopted by Drebin are retrieved from both dexcode and manifest. We provide an overview of MaMaDroid and Drebin below.

\subsection{MaMaDroid}

MaMaDroid extracts features from the CFG of an application. It uses the sequence of abstracted API calls rather than the frequency or presence of certain APIs, aiming at capturing the behavioural model of the mobile application. MaMaDroid operates in two modes, namely \texttt{family} mode and \texttt{package} mode. API calls will be abstracted to either family level or package level according to their mode. For instance, the API call sendTextMessage() is abstracted as:
$$\underbrace{\underbrace{\overbrace{\mathrm{android}}^{\mathrm{family}}\mathrm{.telephony}}_{\mathrm{package}}\mathrm{.SmsManager:void\: sendTextMessage()}}_{\mathrm{API\: call}}$$
Family mode is more lightweight, while package mode is more fine-grained. We demonstrate the results of attacking both.

\begin{figure*}
  \centering
  \includegraphics[width=1\textwidth]{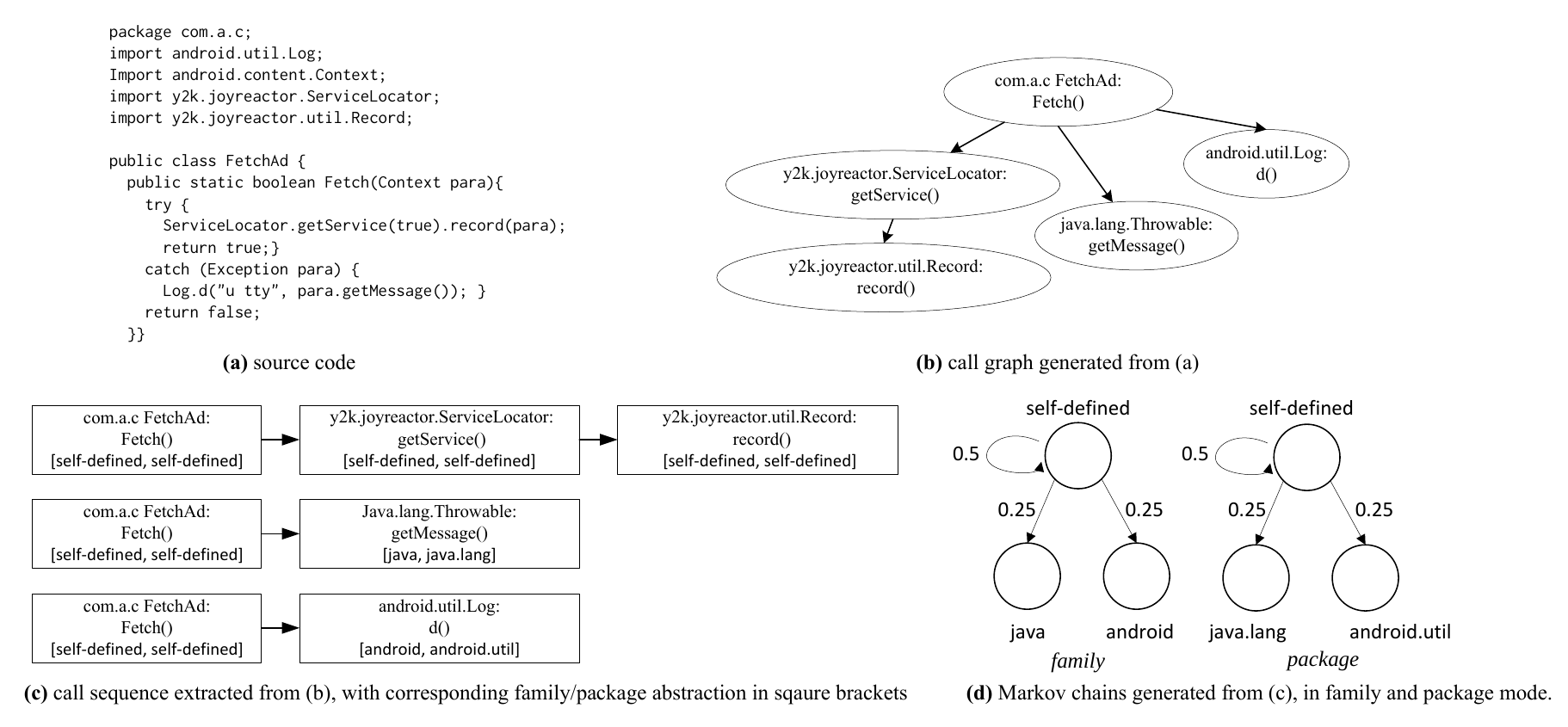}
  \caption{Process of feature extraction in MaMaDroid, from (a) to (d)}
  \label{fig_mamadroid} 
\end{figure*}

MaMaDroid firstly extracts the CFG from each application, and obtains the sequences of API calls. Then, the API calls are abstracted using either of the above-mentioned modes. Finally, MaMaDroid constructs a Markov chain with the transition probabilities between each family or package, used as the feature vector to train a machine learning classifier. Fig. \ref{fig_mamadroid} illustrates the feature extraction process in MaMaDroid. Sub-graph (a) is a code snippet that has been decompiled from a malicious application; sub-graph (b) shows the call graph extracted from the source code; sub-graph (c) is the abstracted call graph generated from (b); and finally, sub-graph (d) presents the Markov chain generated based on (c).  

MaMaDroid recognises nine families and 338 packages from official Android documentation. Packages defined by application developer and obfuscated with identifier mangling, are abstracted as \texttt{self-defined} and \texttt{obfuscated}, respectively. Overall, there are 340 possible packages and 11 families.

Given the extracted features, MaMaDroid leverages RF, KNN, and SVM to train the malware detector and test the performance on several datasets (which were collected over different time periods). RF outperforms the other two classifiers, with its F-measure reaching 0.98 and 0.99 in the family and package modes, respectively.

\subsection{Drebin}
\label{SubS:target_systems_drebin}

Drebin is an on-device lightweight Android malware detector. Drebin extracts features from both the manifest and the disassembled dexcode through a linear sweep over the manifest file and the disassembled smali files of the application. The features such as permissions, activities, and API calls are presented as strings. Eight sets of features are retrieved, as listed in Table \ref{drebin_feature}.

\begin{table}[htbp]
\centering
\caption{Overview of Drebin feature set}
\begin{tabular}{c|l}
\toprule
\multicolumn{2}{c}{Drebin feature sets}                                 \\
\midrule
\multirow{4}{*}{manifest}& $S_1$  Hardware components \\
& $S_2$  Requested permissions \\
& $S_3$  App components        \\
& $S_4$  Filtered intents      \\
\midrule                                             
\multicolumn{1}{l|}{\multirow{4}{*}{dexcode}} & $S_5$  Restricted API calls  \\
\multicolumn{1}{l|}{}& $S_6$  Used permissions      \\
\multicolumn{1}{l|}{}& $S_7$  Suspicious API calls  \\
\multicolumn{1}{l|}{}& $S_8$  Network addresses     \\
\bottomrule
\end{tabular}
\label{drebin_feature}
\end{table}

The extracted features are put into a multidimensional vector (\textit{S}) to create a $\lvert S\rvert$-D space, in which we can have 0 or 1 value along each dimension, indicating the presence or absence of the corresponding feature. The following shows an example of the feature vector $\varphi (x)$ of a malicious application that sends premium SMS messages and thus requests certain permissions and hardware components. 

$$\varphi (x) \mapsto 
\begin{pmatrix}
\ \texttt{...}\ \\ 
\ \texttt{0}\ \\ 
\ \texttt{1}\ \\ 
\ \texttt{...}\ \\ 
\ \texttt{1}\ \\ 
\ \texttt{0}\ \\ 
\ \texttt{...}\ \\ 
\end{pmatrix}
\begin{matrix}
\ \texttt{...}\\ 
\texttt{permission.SEND\_SMS}\\ 
\texttt{permission.RECORD\_AUDIO}\\ 
\ \texttt{...}\\ 
\texttt{hardware.camera}\\ 
\texttt{hardware.telephony}\\ 
\ \texttt{...}\\ 
\end{matrix}
$$
After the features being retrieved, Drebin learns a linear SVM classifier to discriminate between benign and malicious applications. The classification performance on Drebin was evaluated on a dataset consisting 5,560 malware samples and 123,453 benign applications, which are collected between August 2010 and October 2012.  

\subsection{Attack Scenarios}
\label{attack_scenarios}
The knowledge of the target system obtained by the adversary may vary in different situations. This includes the feature set, the training set, the classification algorithm as well as the parameters. We argue that in the real world, it is not likely for the adversary to have full knowledge of the classification algorithm used in the target detector. However, the adversary can probe the detector through feeding desired inputs and getting the corresponding outputs.

Knowing the feature set, as a baseline assumption for attacking learning systems, has been widely adopted in similar works in this field \cite{biggio2013evasion, chen2018automated, grosse2017adversarial, laskov2014practical}. Therefore,
in this paper, we consider the following four situations in our attack: 1) \textit{Scenario F}: the adversary only knows the feature set; 2) \textit{Scenario FB}: The adversary knows the feature set only, and can query the target detector as a black box; 3) \textit{Scenario FT}: The adversary knows both the feature set and training set, but cannot query the target detector; and 4) \textit{Scenario FTB}: The adversary knows both the feature set and the training set, and can also query the target system as a black box.
Note that in the scenarios that allows querying the target system as a black-box (\textit{i.e.}, scenario FTB and FB), the only information that the adversary can get is the predicted label from the black-box oracle when given an input.
Also note that in the scenarios of having access to the training set (\textit{i.e.}, scenario FTB and FT), the adversary can only have a copy of the training set, but he/she cannot inject new samples or modify the existing ones in the training set. 

\section{Attack on MaMaDroid}
\label{sec:attack_on_mamadroid}
\subsection{Attack Algorithm}
\label{attacking_algorithm_mama}
We introduce an evasion attack on MaMaDroid in this section. The purpose is to make a piece of malware evasive with minimal API call injections into its original smali code. We assume that we only have black-box access to the target (MaMaDroid) detector. In other words, we can get output from MaMaDroid by feeding input, but we do not know how it processes internally. There are two considerations for the features used in MaMaDroid. First, because the features are actually the state transition probabilities of the call graph, the probabilities of the transitions departing from the same node in the call graph will increment up to 1. Second, the feature value should be bounded between 0 and 1. We will address these considerations in our algorithms.

We employ two adversarial example crafting algorithms that have been widely adopted to generate evasive malware examples. To study a more effective way of attacking, we craft adversarial example by either optimising an adversarial objective function (\textit{i.e.}, refer as C\&W), or perturbing influential features based on the indicative forward derivatives (\textit{i.e.}, refer as JSMA). C\&W and JSMA are originally developed for crafting adversarial image examples, which has continuous pixel values as the features. In our case, we are going to calculate the perturbation based on the number of  API calls, which are discrete. Therefore, we need to refine plain C\&W and JSMA algorithms to cater our needs. We construct a neural network $F$ as a substitute to launch the attack. In the malware detection case, $F$ is a binary classifier which has a 2D output. Let the input features of the original malware form an $n$ dimensional vector, denoted as $X$.

\subsubsection{Refined C\&W}
\label{SubSubcw}
C\&W crafts adversarial malware with tunable attack confidence while optimising the distortion on the original malwale features. We modify C\&W to search for an adversarial malware sample through optimising an objective function with the following constraints:

\begin{equation}
    \begin{array}{l}
        \min_{\delta}{||\delta||_{2}^{2}+c\cdot f(X+\delta)}\\
        \\
        s.t.\ X+\delta\in [0,1]^{n},\\
        \\
        and\ ||X_g+\delta_g||_{1}=1, g\in{1...k}.
    \end{array}
\end{equation}
Here, $\delta$ is the perturbation to be optimised and $c$ is a constant to balance the two terms in the objective function. We use line-search to determine the value of $c$. The first term in the objective function minimises the $l_2$ distortion on the original features, which means the change on the MaMaDroid feature should be small enough to limit the amount of API calls we insert into the smali code. The second term is a specially designed adversarial loss function $f$. Suppose $t$ is the ground-truth class of the current malware example $X$. Our goal is to make $X$ be incorrectly classified into the other classes (in our case, the benign class). Thus, $f$ takes the following format:

\begin{equation}
    f(X)=max(Z(X)_{t}-max\{Z(X)_{i}:i\ne t\},-\kappa)
\end{equation}

in which $Z(X)$ is the pre-softmax output from the substitute $F$, $\kappa$ is a hyper-parameter which can adjust the attack confidence and $f$ will maximise the loss between the output of current model and the ground-truth. To address the aforementioned considerations, we apply two constraints in the optimisation. First, each feature after perturbation should be between 0 and 1. Second, the $l_1$ norm of the features in each family/package group should be equal to 1. The objective function is optimised with AdaGrad \cite{duchi2011adaptive}. The feature values are iteratively updated until being misclassified. We use either the substitute model (in scenario F and FT), or the MaMaDroid oracle (in scenario FB and FTB), which we refer as the \textit{pilot classifier}, to determine whether a sample is misclassified. 

Since the current feature $X$ is a set of probabilities, to make the perturbation viable during the code injection into the original smali code, we change the optimisation variable from $\delta_i$ on $X$ (the perturbation on the probabilities) to $\omega$ on $A$ (the perturbation on the number of API calls). For the perturbation on the $i$-th feature in group $g$, we have:

\begin{equation}
    \delta^g_i = \frac{a^g_i+\omega^g_i}{a^g+\omega^g}-\frac{a^g_i}{a^g}.
\end{equation}
wherein $\omega^g=\sum_i{\omega^g_i}$ and $a^g_i$ is the number of API calls indicated by the $i$-th feature in the $g$-th group. We change the optimiser from $\delta$ to $\omega$. Accordingly, we change the first term of the adversarial objective function to $||\omega||^2_2$, in order to minimise the total number of code injections.

Deleting codes may jeopardise the functionality of the malware, therefore we only inject code to make adversarial examples. We apply a ReLu function (\textit{i.e.}, $ReLu(\omega)=max(0,\omega)$) to clip $\omega$ to non-negative values after each iteration. As the result, the first constraint (\textit{i.e.}, $\frac{a^g_i+\omega^g_i}{a^g+\omega^g} \in [0,1]$) is automatically satisfied. To satisfy the second constraint (the sum of the feature values in the same group being 1), we normalise $\sum_{i}{\frac{a^g_i+\omega^g_i}{a^g+\omega^g}}$ for each group after each gradient descent round. 

\subsubsection{Refined $JSMA$}
JSMA finds adversarial examples using the forward derivatives of the classifier. JSMA iteratively perturbs important features to determine the Jacobian matrix based on the model input and output features. The method first calculates the Jacobian matrix between the input features $X$ and the outputs from $F$. In the case of MaMaDroid, we want to find the Jacobian between the API call numbers $A$ and the outputs from $F$, given the relationship between API call numbers and the probabilities (\textit{i.e.}, the input features). The Jacobian can be calculated as follows:

\begin{equation}
    J_{F}{(A)} = [ \frac{\partial F(X)}{\partial X} \frac{\partial X}{\partial A} ] = [\frac{\partial F_j{(X)}}{\partial x_i}\frac{\partial x_i}{\partial a_i}]_{i\in{1...n}, j\in{0, 1}}
\end{equation}
wherein $i$ is the index of the input feature and $j$ is the index of the output class labels (in our case it is binary). $x_i$ is the $i$-th feature, $a_i$ is the corresponding $i$-th API call, and $F_j{X}$ is the output of the substitute at the $j$-th class. Suppose $t$ is the ground truth label. To craft an adversarial example, $F_t{(X)}$ should decrease while the outputs of other classes $F_j{(X)}, j\ne t$ are increased.

Based on the calculated Jacobian, we can construct a saliency map $S(A,t)$ to direct the perturbation. The value for feature $i$ in the saliency map can be computed as:

\begin{equation}
    S(A,t)[i] = \begin{cases}
             0, if J_{it}{(A)}>0\ or\ \sum_{j\ne t}{J_{ij}{(A)}<0,}\\
             |J_{it}(A)|(\sum_{j\ne t}{J_ij(A)}), otherwise.
             \end{cases}
\end{equation}

According to the saliency map, we pick one API call ($i$) that has the highest $S(A,t)[i]$ value to perturb during each iteration. The maximum amount of allowed changes is restricted to $\gamma$. The number of the selected API call will be increased by a small amount, represented as $\theta$, in each iteration. The iteration terminates when the sample is misclassified by the \textit{pilot classifier}, or the maximum change number is reached. 

\subsection{APK Manipulation}
\label{apk_modification_mama}

In our study, the development of the APK file modification method was guided by the following design goals: 1) the modified APK will keep its original functionality; and 2) the modification will not involve additional human efforts, \textit{i.e.}, it can be applied automatically via running scripts.

As introduced in section \ref{target_system}, the feature vector that MaMaDroid uses are the transition probabilities between \textbf{states} (either families or packages). Intuitively, the modification approach we apply is to add a certain number of API calls from specific callers to callees into the code to change feature values in the feature space. Since we can obtain the total number of calls that go from any callers to any callees with static analysis, we therefore can calculate how much the feature values will be affected by adding a single call.

The APK manipulation process is designed with two strategies, namely simple manipulation strategy and sophisticated manipulation strategy. The following explains their details and limitations, respectively.

\textbf{Simple manipulation strategy} was motivated by the process that MaMaDroid extracts and calculates its feature values. MaMaDroid extracts all API calls from  \texttt{classes.dex}, and abstracts them as either their families or packages merely based on their root domain in the package names. For instance, The self-defined class \texttt{"MyClass"} in a self-defined package like \texttt{android.os.mypack}, and the system class \texttt{"StorageManager"} in the system package \texttt{android.os.storage}, will both be abstracted as \texttt{android} family or \texttt{android.os} package. By adding such self-defined classes, we are able to mislead the abstraction of API calls in MaMaDroid. 

According to the above observation, we design some  code blocks that can include an arbitrary number of calls from any caller to any callee. The java source code shown below is an example of adding two \texttt{android to android} calls. Arbitrary number of calls can be added by simply invoking \texttt{callee()} multiple times in the \texttt{caller()}.   

\begin{lstlisting}[]
package android.os.mypack

public class Myclass {
    public static void callee() {}
    public static void caller() {
        callee();
        callee();}}
\end{lstlisting}

Our approach proceeds by injecting the required self-defined classes into the source of the target APK, and invoking the corresponding \texttt{caller} methods in the \texttt{onCreate()} method of its entry point activity class. The entry point activity can be located by searching \texttt{"android.intent.action.MAIN"} in the manifest. Since source code cannot be perfectly reverse-engineered to Java, we perform the code insertion on the smali code. As mentioned in Section \ref{apk}, the modified smali codes can be rebuilt to make an APK again. The following listing presents the smali code of the above Java source code (with constructor methods omitted). 
\begin{lstlisting} []
.class public Landroid/os/mypack/Myclass;
.source "Myclass.java"

.method public static callee()V
    .locals 0
    return-void
.end method

.method public static caller()V
    .locals 0
    .line 6
    invoke-static {}, Landroid/os/mypack/Myclass;->callee()V
    return-void
.end method
\end{lstlisting}

The described modification process can add an arbitrary number of calls from any callers to any callees, by simply runing an automated script. It also ensures that the process will not affect the functionality of the original application. However, it modifies the CFG that MaMaDroid extracted from the APK, and consequently modifies its feature values.

Simple manipulation takes advantage of the design flaw in the feature abstraction process in MaMaDroid, thus can possibly be defended by implementing white-list filter, which filters out the API calls that are not in a standard Android SDK when processing API abstraction (which is not implemented in MaMaDroid).  

\textbf{Sophisticated manipulation strategy} is designed to bypass the white-list filter, in which system provided non-functional API calls are inserted into the smali code. For instance, invoking a \texttt{Log.d()} method in the \texttt{onCreate()} method of the entry \texttt{activity} class (\textit{e.g.},  \texttt{com.my.project.MainActivity}), will result in adding one \textit{self-defined to android} call in the family mode, or one \textit{self-defined to android.util} call in the package mode. Since the calls that we inserted are in the \textit{activity} class of the project, it is abstracted to \texttt{self-defined} or \texttt{obfuscated} according to the abstraction rule of MaMaDroid. Therefore, with sophisticated manipulation, calls only originated from self-defined or obfuscated family/package can be inserted. Such limitation slightly decreased the evasion rate from 99\% to 93\% in our family mode experiment. An example of added smali code for a \texttt{log.d()} method is presented as follows.

\begin{lstlisting} []
const-string p0, ""
const-string p1, ""

.line 13
invoke-static {p0, p1}, 
     Landroid/util/Log;->d(Ljava/lang/String;
     Ljava/lang/String;)I
\end{lstlisting}

\begin{figure*}
  \centering
  \includegraphics[width=1\textwidth]{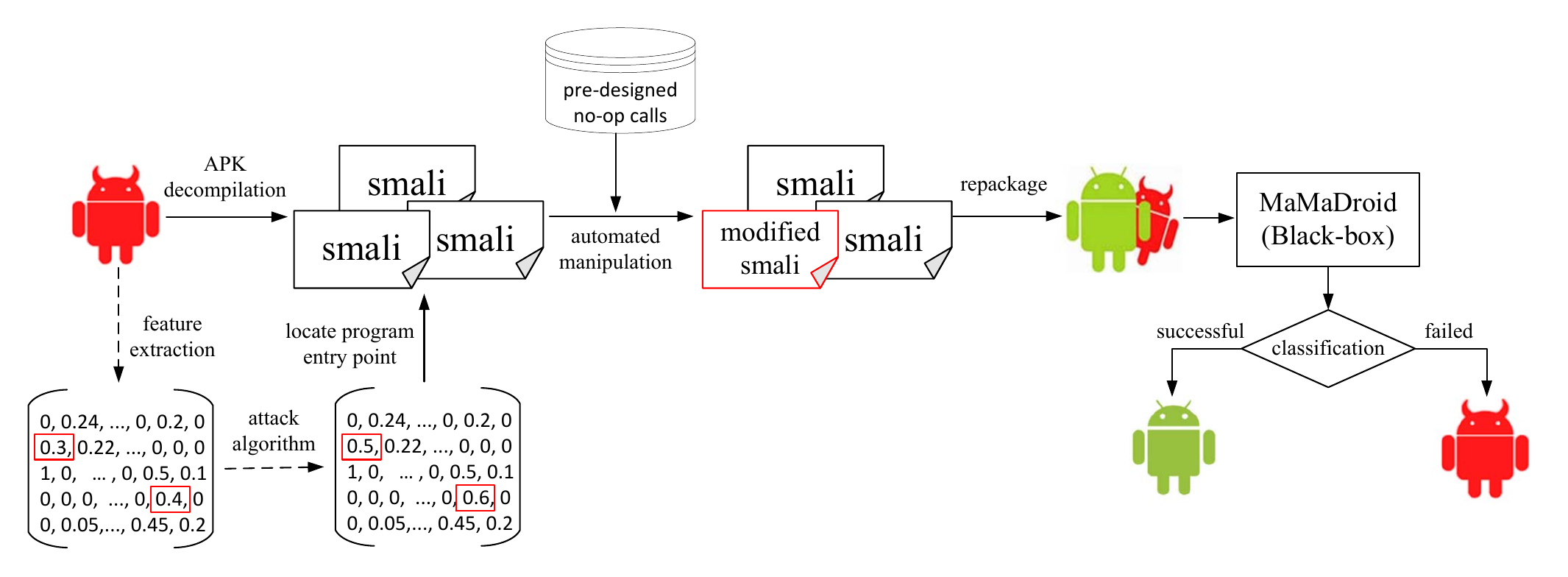}
  \caption{The attack process: the dashed lines show the process of our attack algorithm, and the solid lines illustrate our APK manipulation procedure.}
  \label{fig_attack_process} 
\end{figure*}

We developed a script to automatically perform the code insertion process. We firstly prepared the above described \textit{no-op} code blocks from each caller to each callee. These code block are independent to the application, thus can be repeatedly used in the attack. The number of calls to be inserted from specific callers to callees were calculated by our attack algorithms described in Section \ref{attacking_algorithm_mama}. Then, we used regular expression to locate the \texttt{onCreate()} method in the smali code of the entry point \textit{activity} class, and add any necessary code blocks to the end of the \texttt{onCreate()} method. Fig. \ref{fig_attack_process} demonstrates the attack process, in which the dashed lines show the process of our attack algorithm, and the solid lines illustrate our APK manipulation procedure.  

\subsection{Experiment Settings}
\label{evaluation_mama}

The experiments to be presented in the following two subsections evaluate the effectiveness of crafted adversarial examples. More specifically, we are going to answer the following two questions: 1) can the modified malware sample effectively evade from the target detector? and 2) can the modification be easily applied to the original APK? For the convenience of experiments, we built MaMaDroid based on the source code that the authors published online\footnote{https://bitbucket.org/gianluca\_students/mamadroid\_code}.

\subsubsection{Dataset}

To evaluate the performance of the crafted adversarial examples, we use the same datasets that have been used in MaMaDroid. First, the set of benign applications consists of 5,879 benign applications collected by PlayDrone \cite{viennot2014measurement} in 2014 (denoted by \textit{oldbenign} in \cite{mariconti2016mamadroid}). The set of malware includes 5,560 samples that were initially used in Drebin \cite{arp2014drebin} and collected between 2010 and 2012 (denoted by \textit{drebin} in \cite{mariconti2016mamadroid}). The original experiments reported in \cite{mariconti2016mamadroid} also tested several combinations of other {old} and {new} datasets collected over years to evaluate the robustness of their approach. Using only one set of data does not affect our research target, \textit{i.e.}, to craft adversarial example that can fool and evade the malware detector. The classification results on the chosen datasets are promising, of which the F-measures reach 0.88 and 0.96, in the family and package modes, respectively. Our work is to generate malware samples for evading the detection, therefore, our test set obtains only malware samples. We carefully prepare the test set by manually checking that every sample can be installed and launched on an Android smart phone. We randomly select 1,000 qualified malware samples to form the test set, leaving the rest of the malware samples, together with the benign application samples to be the training set.

As discussed in Section \ref{attack_scenarios}, to simulate the scenarios where the original training dataset of the target detector is unknown to the adversary (\textit{i.e.}, Scenario F and FB), we collected a set of malware samples and another set of benign applications from VirusShare\footnote{https://virusshare.com} and APKPure\footnote{https://apkpure.com}, respectively. VirusShare dataset consists of 24,317 malware samples collected between May 2013 to March 2014, while APKPure dataset consists of 10,000 applications we crawled from its website on January 2018. The applications from APKPure are submitted to VirusTotal to examine their benignity. We discard the samples that are reported by at least one anti-virus engine as malicious. Finally, the APKPure dataset contains 9,664 application.
We randomly selected 4,560 malware samples and 5,879 benign applications from VirusShare and APKPure datasets, respectively, to form the surrogat dataset (to eliminate the influence caused by different number of training samples in the original and surrogate datasets).
In the FT and FTB scenarios, we use the \texttt{original} dataset to train the target detector, as well as our attack algorithm; while in the F and FB scenarios, we use the \texttt{original} dataset to train the target detector and the \texttt{surrogate} dataset to train the attack algorithm.

\subsubsection{Experiment Work Flow}

Given a malicious APK as the input, we firstly decompiled it with apktool, and constructed its feature vector. The attack algorithm then optimised the perturbations to be added to the feature vector, \textit{i.e.}, the number of calls added from each caller to callee. Then, corresponding pre-designed code blocks were inserted into the smali files, which were then recompiled into a new APK. The manipulated APK was submitted to MaMaDroid oracle to get the classification result. The attack was declared successful if the modified APK was labelled as benign. This process makes sure that our attack method not only changes the feature vector, but also effectively modifies the APK. We additionally verified that all the modified APKs can be successfully installed and launched on an Android smartphone. It was difficult to verify whether the functionality was affected or not. However, we presume that since the calls we added were non-functional, they will not have changed the functionality of original APK. 

As we have explained before, we run experiments in four deliberate scenarios (refer to Section \ref{attack_scenarios}). The details of the settings for each scenario are listed in Table \ref{scenario_overview2}. In the experiments, we train a substitute model to approximate MaMaDroid by using AdaGrad. Accordingly, a multi-layer perceptron (MLP) model is employed. The model contains 2 fully connected hidden layers, each has 128 nodes. Each training batch contains 256 samples and the substitute model is trained for 100 epochs. In addition, we introduce dropout after each hidden layer to prevent overfitting problem in the experiments. We set the dropout rate to 0.5. Note that MaMaDroid trained with the original dataset is used as benchmark for evaluation, and we only require black-box access to the pilot classifier (refer the definition to Section \ref{attacking_algorithm_mama}).

\begin{table}[b]
\centering
\caption{Attack Scenarios}
\begin{tabular}{ccc}
\toprule
Scenario & Pilot Classifier & Training Set \\
\midrule
F        & Substitute       & Surrogate    \\
FT       & Substitute       & Original     \\
FB       & MaMaDroid        & Surrogate    \\
FTB      & MaMaDroid        & Original     \\
\bottomrule
\end{tabular}
\label{scenario_overview2}
\end{table}

\subsection{Experiment Results}
In \cite{mariconti2016mamadroid}, MaMaDroid's performance was examined on three different machine learning classifiers. They are RF, SVM, and K-Nearest Neighbour (KNN). To be consistent with the experiments in \cite{mariconti2016mamadroid}, we also evaluate our proposed method on these classifiers, respectively.
In addition, to investigate the robustness of Deep Neural Networks (DNN) in malware detection, we leverage the features of MaMaDroid to train a DNN-based detector. Specifically, our DNN-based detector consists of five hidden layers, each with 128, 64, 64, 64, 64 neurons, respectively. The F-measures in family and package modes are 0.92 and 0.95, respectively, which is comparable to the state-of-the-art. The proposed attack methods are also evaluated on the DNN-based detector.

The effectiveness of the crafted adversarial examples is evaluated in terms of evasion rate and distortion. \textbf{Evasion rate} is defined as the ratio of malware samples that are misclassified as benign, to the total number of malware samples in the testing set. \textbf{Distortion} is defined as the number of API calls added to the smali code for each malware sample.

\subsubsection{Overall results} 
\begin{figure*}
  \centering
  \includegraphics[width=1\textwidth]{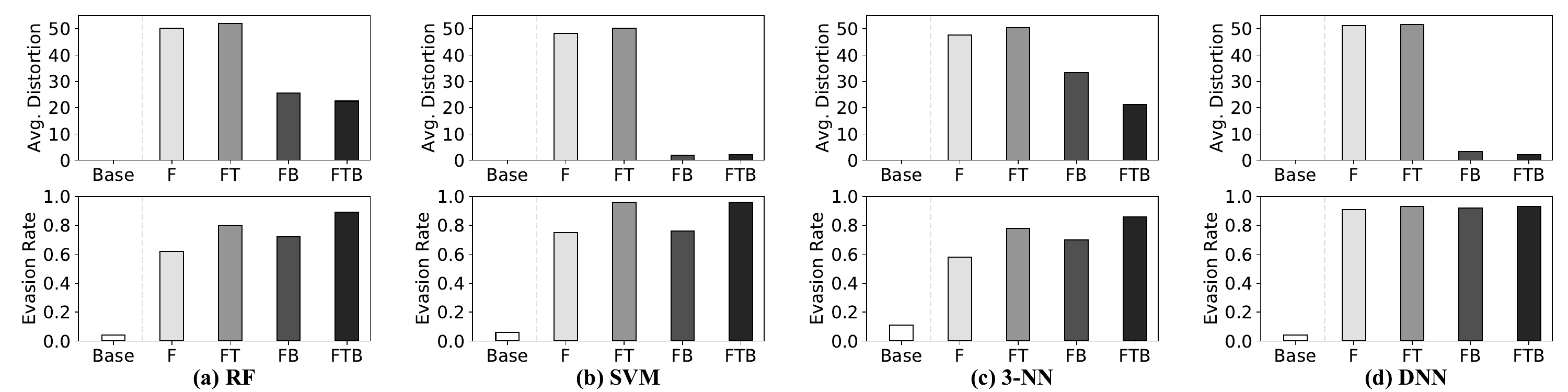}
  \caption{The evasion rate and average distortion of adversarial examples generated by \textit{JSMA} in the \textit{family} mode }
  \label{fig_result_family_jsma} 
\end{figure*}

\begin{figure*}
  \centering
  \includegraphics[width=1\textwidth]{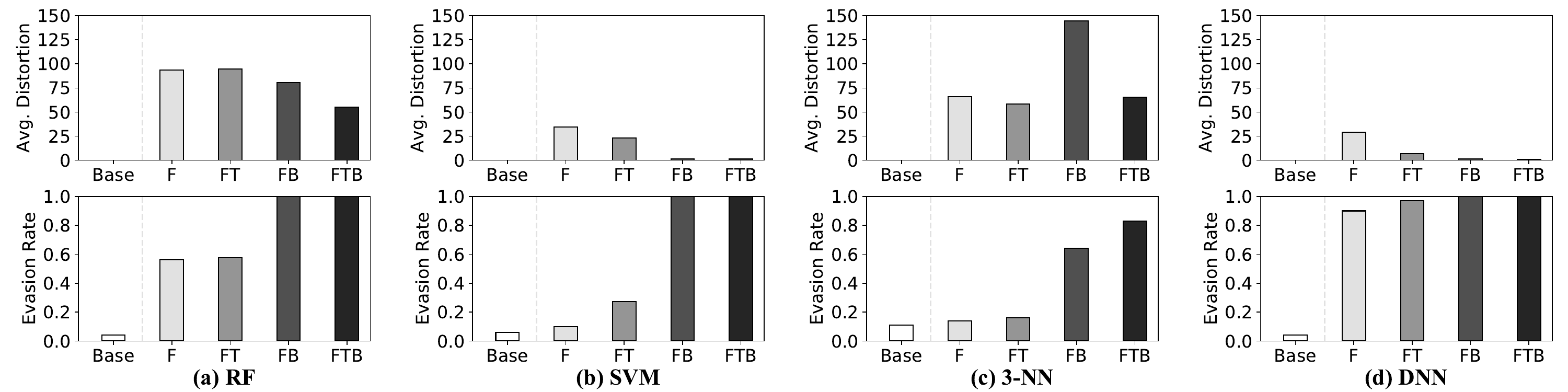}
  \caption{The evasion rate and average distortion of adversarial examples generated by \textit{C\&W} in the \textit{family} mode.}
  \label{fig_result_family_cw} 
\end{figure*}

\begin{figure*}
  \centering
  \includegraphics[width=1\textwidth]{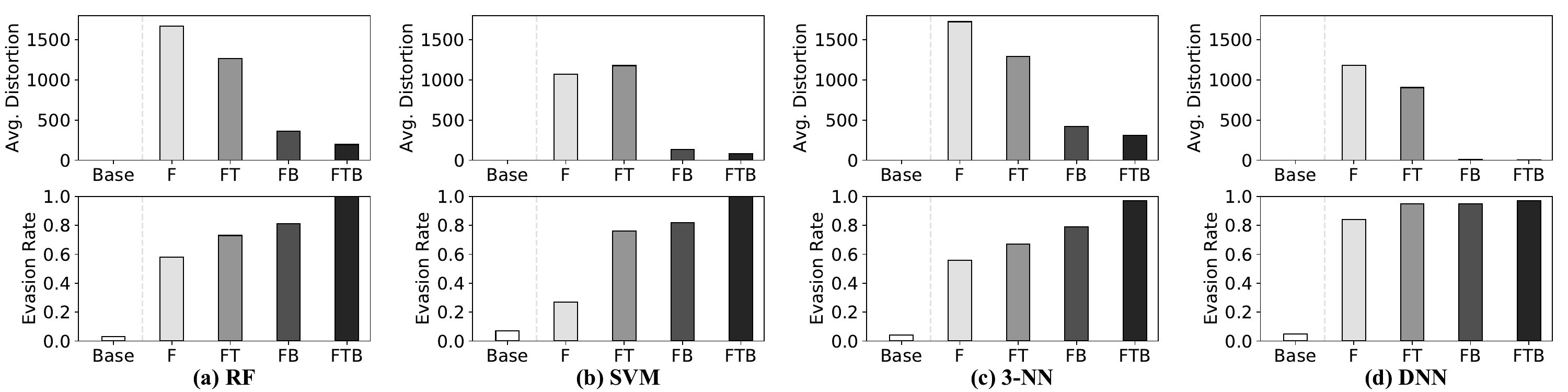}
  \caption{The evasion rate and average distortion of adversarial examples generated by \textit{JSMA} in the \textit{package} mode.}
  \label{fig_result_package_jsma} 
\end{figure*}

\begin{figure*}
  \centering
  \includegraphics[width=1\textwidth]{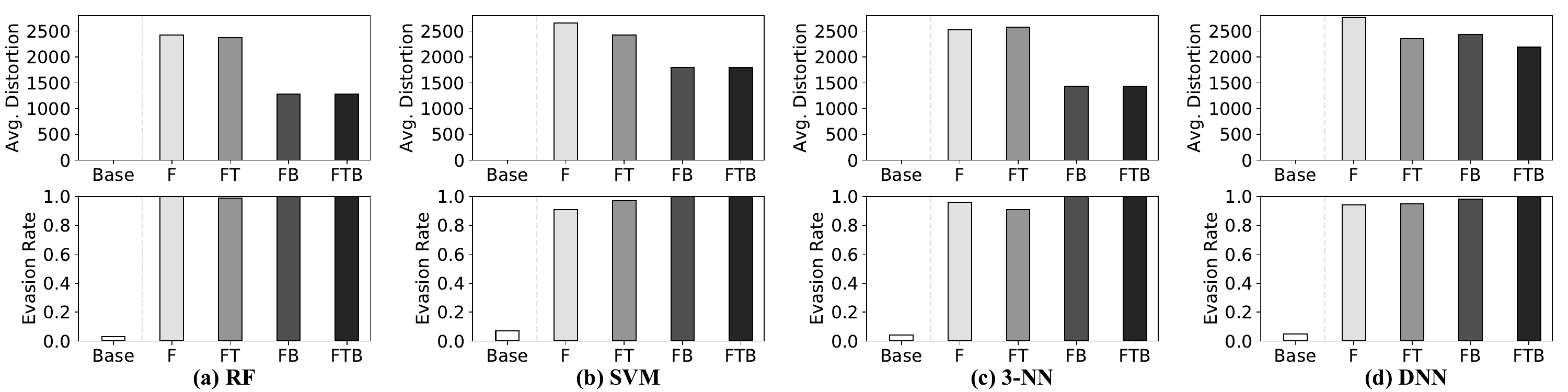}
  \caption{The evasion rate and average distortion of adversarial examples generated by \textit{C\&W} in the \textit{package} mode.}
  \label{fig_result_package_cw} 
\end{figure*}

The overall results of our attack are presented in Fig. \ref{fig_result_family_jsma}-\ref{fig_result_package_cw}. Specifically, Fig. \ref{fig_result_family_jsma} and Fig. \ref{fig_result_family_cw} present the attack results of family mode using two attack algorithms, while Fig. \ref{fig_result_package_jsma} and Fig. \ref{fig_result_package_cw} demonstrate the results of package mode. We applied the attack on aforementioned four machine learning algorithms (sub-figures (a)-(d)), under four real world scenarios (x-axes) as discussed in Section \ref{attack_scenarios}. Simple manipulation strategy is applied in this experiment, while sophisticated manipulation strategy is evaluated in Sub-section (4). The evasion rate before attack is also reported and acted as a baseline. The evasion rate as well as the average distortion for each sample is reported. 
The results indicate that the proposed attack methods effectively evaded MaMaDroid in most of the real world scenarios. For instance, the evasion rate on RF increased from 4\% (before attack) to 56\%-99\% (after attack) in the family mode, and from 3\% to 58\%-99\% in the package mode, depending on the scenario and attack algorithm. It is worth to note that in scenario FTB, where adversary gains most knowledge of MaMaDroid, the evasion rate (C\&W) reaches 100\% in RF, 100\% in SVM, 83\% in 3-NN, and 100\% in DNN, with average 55, 2, 65, and 1 API calls added to each malware samples for these algorithms, respectively. Even when the adversary only knows the feature set (\textit{i.e.}, scenario F), the evasion rates with JSMA reach 62\%, 75\%, 58\%, and 91\%, in the above mentioned algorithms, respectively. 

\begin{figure*}
  \includegraphics[width=1\textwidth]{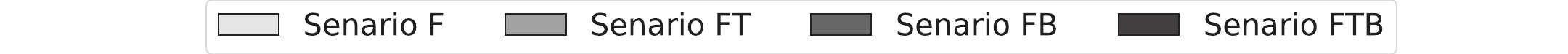}
\end{figure*}

\subsubsection{Evaluation results by scenarios} 

An important observation is the improvement of attack effectiveness with the increase of adversary's knowledge of the target system. 
While different level of knowledge obtained by adversary affects the evasion rate in both algorithms, the impact on each factor is different. As demonstrated in Fig. \ref{fig_result_family_cw} , in the scenarios which black-box access to MaMaDroid oracle is acquired (\textit{i.e.}, FB and FTB), the evasion rate of C\&W in all four algorithms are significantly higher than the evasion rate in the scenarios which black-box access is not granted (\textit{i.e.}, F and FT). In the meanwhile, the possession of training set (F \textit{versus} FT, FB \textit{versus} FTB) has little impact on the evasion rate. However, the evasion rate in JSMA (refer to Fig. \ref{fig_result_family_jsma}) are to the contrast. The possession of training set influenced the evasion rate significantly, while the access to black-box model is less important.  

\subsubsection{Evaluation results by operation modes} As introduced in Section \ref{target_system}, MaMaDroid runs in either the family mode or the package mode. Family mode is more lightweight, while package mode is more fine-grained. The original classification performance in the package mode is slightly better than that in the family mode, with the original (baseline) evasion rate falls in the range of 1\%-6\% on various algorithms (compared with 4\%-11\% in the family mode). The results of the experiment indicate that the attack is more effective in the package mode than in the family mode, in terms of evasion rate. For instance, when attacking using JSMA, the evasion rate in the package mode with RF reaches 100\% in scenario FTB (Fig. \ref{fig_result_package_jsma}(a)), while it is 89\% in the family mode in the same scenario (Fig. \ref{fig_result_family_jsma}(a)). However, the average distortion of the adversarial example in the package mode is significantly higher than in the family mode. In average, 17 API calls need to be added in each application in the family mode, while this number increased to 257 in the package mode. The results disclose that while using more fine-grained features slightly enhance the classification accuracy, it's resistance to our attack is significantly higher than using highly abstracted features (\textit{i.e.}, family mode), considering that more than 15 times of number of calls need to be inserted for a successful evasion. 

\subsubsection{Evaluation results by manipulation strategy} 
\begin{figure}[t]
  \centering
  \includegraphics[width=1\columnwidth]{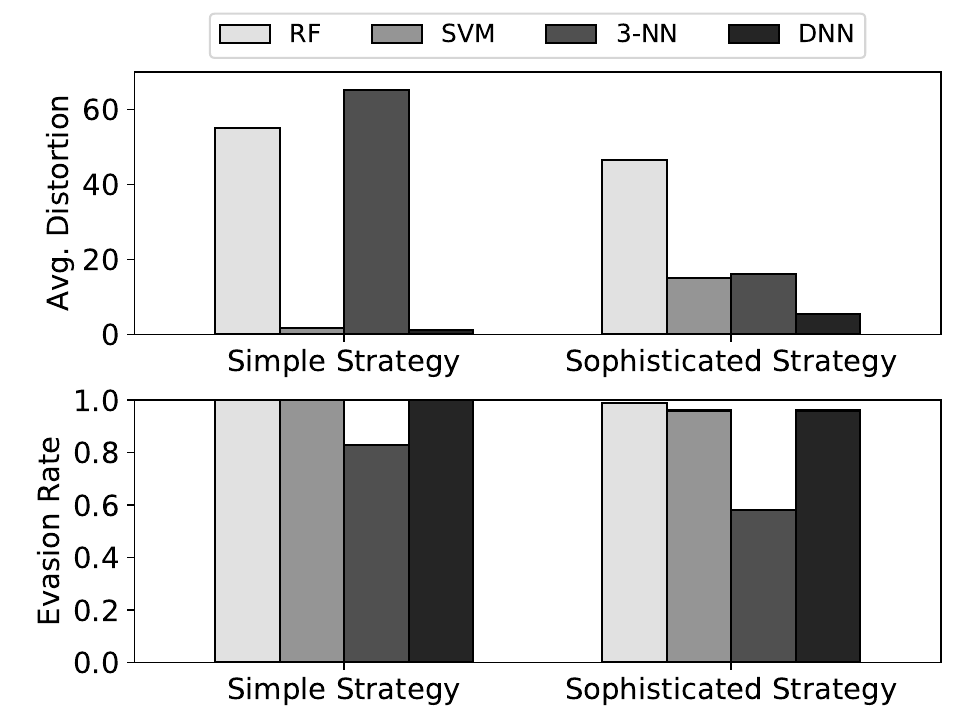}
  \caption{Comparison of applying simple manipulation strategy and sophisticated manipulation strategy in the family mode by C\&W}
  \label{fig_result_mama_each_group} 
\end{figure}

As presented in Section \ref{apk_modification_mama}, two strategies can be applied to the proposed APK manipulation method. In simple manipulation strategy, API calls originated from any caller can be inserted into the smali code; while in sophisticated manipulation strategy, only API calls originated from \texttt{android}, \texttt{google}, \texttt{self-defined} and \texttt{obfuscated} can be added. Thus, we examine the feasibility of the sophisticated manipulation strategy, by restricting that only the values of the calls originated from aforementioned families can be modified in the feature space. 
Fig. \ref{fig_result_mama_each_group} presents the evasion rates and the corresponding average distortions of applying simple manipulation strategy and sophisticated manipulation strategy in scenario FTB, respectively. 
In the simple manipulation strategy, the evasion rate are 100\%, 100\%, 83\%, and 100\%, in RF, SVM, 3-NN, and DNN, respectively, with 55, 2, 65, and 9 API calls in average to be added; while in the sophisticated manipulation strategy, the evasion rate slightly decreased to 99\%, 96\%, 58\%, and 100\%, respectively. The number of API calls to be injected in the sophisticated manipulation strategy are in average 46, 14, 16, and 9, respectively. The results demonstrate that sophisticated manipulation strategy can also achieve a high evasion rate with only a small number of API calls to be injected into the APK. 

\begin{figure}[h]
  \centering
  \includegraphics[width=1\columnwidth]{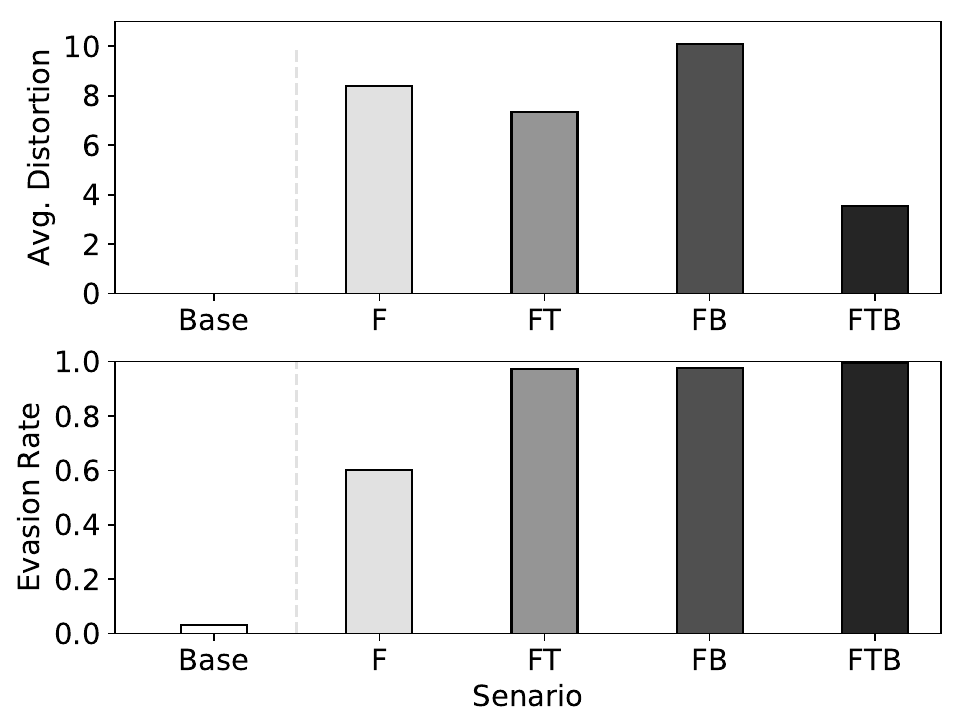}
  \caption{The average distortion and evasion rate of adversarial example generated by \textit{JSMA} on Drebin.}
  \label{fig_result_drebin_overall} 
\end{figure}

\section{Attack on Drebin}
\label{attacking_drebin}

\subsection{Attack Algorithm}
We adopt the Jacobian-based attack to craft an adversarial example for Drebin, since the features of Drebin are binary. $JSMA$  perturbs a feature from 0 to 1 in each iteration. Regarding the Jacobian for Drebin, we calculate it based on the following formula:

\begin{equation}
    J_{F}{(X)} = [ \frac{\partial F(X)}{\partial X} ] = [\frac{\partial F_j{(X)}}{\partial x_i}]_{i\in{1...n}, j\in{0, 1}}
\end{equation}
wherein $X$ is the binary feature vector for Drebin and $i$ is the classification result (\textit{i.e.}, malware if $i=1$). Based on the Jacobian matrix, we select the most influential feature to perturb in each iteration. In other words, we perturb the $i$-th feature for which $i=arg\  max_{i\in{1...n},x_i=0}F_0{(x_i)}$.
We change the selected one feature from 0 to 1 in each iteration, until the example is misclassified, or we reach the maximum  amount of allowed change (\textit{i.e.}, $\gamma$).

\subsection{APK Manipulation}
\label{SubS:drebin_manipulation}
Drebin extracts features from both \texttt{manifest} and \texttt{dexcode}. Different from previous work that only modifies the features in manifest \cite{grosse2017adversarial}, we analyse the capability of modifying the features obtained from the \texttt{dexcode}. 

As explained in Section \ref{SubS:target_systems_drebin}, Drebin retrieves features by applying a linear scan on related source files (\texttt{AndroidManifest.xml} and smali files), which only searches for the presence of particular strings (\textit{e.g.}, name of API calls), rather than examining whether the calls are executed. Therefore, our strategy is to add code containing the required features but never being invoked or executed. The listing below presents an example of adding a ``\texttt{suspicious API: getSystemService()}'' feature to the smali code.

\begin{lstlisting}[]
.method private addSuspiciousApiFeature()V
    .locals 1
    const-string v0, "phone"
    .line 17
    invoke-virtual {p0, v0},
        La/test/com/myapp/MainActivity;->
        getSystemService(Ljava/lang/String;)
        Ljava/lang/Object;
    move-result-object v0
    check-cast v0, Landroid/telephony/TelephonyManager;
    return-void
.end method
\end{lstlisting}

\subsection{Experiments \& Evaluations}
We present our attack performance on Drebin by reporting the evasion rate and the average distortion in different real world scenarios. 
Dataset described in Section \ref{evaluation_mama} is used in the experiments.

Fig. \ref{fig_result_drebin_overall} reports the result of our proposed attack. In scenario FTB, where the adversary gets most knowledge of Drebin (\textit{i.e.}, the feature set, the training set, and output from Drebin oracle), 99\% of malware samples in the testing set are misclassified after the attack, with average 3.5 features to be added in each sample. While in scenario F, where the adversary obtains least knowledge of Drebin (\textit{i.e.}, only the feature set), 60\% adversarial malware examples can evade from detection.

Table \ref{drebin_modified_feature} presents the average number of features inserted into each malware sample, from which we observe that the most added features are in the sets of \textit{restricted API calls} and \textit{suspicious API calls}. 

\begin{table}[htbp]
\centering
\caption{Number of features added in each set}
\begin{tabular}{c|c|c}
\toprule
Source File  & Feature Sets                     & Avg. Number Added \\
\midrule
\multirow{4}{*}{dexcode} & $S_5$ Restricted API calls & 2.17 \\
                         & $S_6$ Used permissions     & 0.1  \\
                         & $S_7$ Suspicious API calls & 1.21 \\
                         & $S_8$ Network addresses    & 0.02 \\
\bottomrule
\end{tabular}
\label{drebin_modified_feature}
\end{table}

\section{Discussion} 
\label{discussion}

\subsection{Comparison with Existing works}

We compare our attack method with another two works in evading machine learning based malware detection. Chen et al. \cite{chen2018automated} proposed to poison the training dataset to mislead machine learning detectors. Grosse et al. \cite{grosse2017adversarial} proposed a white-box attack against deep learning based malware detection models. 
Both of these works require the access to the feature set, the training set and the machine learning model, therefore we compare our results of scenario FTB with them, which requires the same knowledge of the target model. However, the comparison works make additional assumptions. \cite{chen2018automated} further assumes that the adversary is capable of injecting tainted samples into the training set, and \cite{grosse2017adversarial} only considers the situation that the adversary knows the detailed structure and parameters of the targeting machine learning model (\textit{i.e.}, white-box). These assumptions are more restrictive that are unlikely to happen in real world scenarios.

Table \ref{tab:comparison} presents the evasion rate of our methods and the methods proposed in \cite{chen2018automated} and \cite{grosse2017adversarial}.
The results show that our methods outperform the compared methods in terms of evasion rate. 

\begin{table}[htbp]
\centering
\caption{Comparison with existing works (Evasion Rate)}
\begin{tabular}{cccccc}
\toprule
\multirow{2}{*}{} & \multicolumn{4}{c}{MaMaDroid} & \multirow{2}{*}{Drebin} \\                        \cmidrule(lr){2-5} 
                  & RF    & SVM     & 3-NN  & DNN   &     \\ \toprule
Our JSMA Attack & \textbf{89\%}  & \textbf{96\%}    & \textbf{86\%} & \textbf{93\%}  & \textbf{99.4\%}  \\ 
Our C\&W Attack & \textbf{100\%} & \textbf{100\%}   & \textbf{83\%} & \textbf{100\%} & \NA      \\ 
Chen et al. \cite{chen2018automated} & \NA    & 68.95\% & \NA   & \NA    & 75.2\%  \\ 
Grosse et al. \cite{grosse2017adversarial} & \NA    & \NA      & \NA   & \NA    & 69.35\% \\ 
\bottomrule
\end{tabular}
\label{tab:comparison}
\end{table}

\subsection{Applicability of Our Attack}
A great number of machine learning based Android malware detection techniques have been proposed in the past few years. The main differences and key contributions of these techniques are the features they extracted to profile the malware samples. As we could not demonstrate the effectiveness of the proposed method on every machine learning detector, we therefore selected two typical detectors, Drebin and MaMaDroid that use syntactic and semantic features, respectively. In prior works, they have been selected as baseline methods to evaluate the performance of adversarial attacks, e.g., \cite{chen2018automated, grosse2017adversarial}. 

Our proposed attack framework can be applied in most machine learning based detectors, which extract features from either manifest or bytecode of an application. We just need to refine the attack algorithm according to the constraint and inter-dependency of the features used in the target detector, just as what we have done for attacking MaMaDroid and Drebin.

\subsection{Artifacts in Our Attack}
It could be argued that adding a certain number of dummy calls or no-op APIs, such as logging output and reading files, introduces artifacts into the APK, which may make the application look suspicious. To investigate whether our attack will introduce such side effect to the original APK, we observe the prevalence of such no-op API calls (\textit{e.g.} android.util.log) in the applications in the wild. According to our observation on the experiment dataset, we find that it is normal for an application, either benign or malicious, to have a certain number of no-op API calls. For example, 17.9\% benign applications and 16.3\% malware samples in the dataset have more than 100 \texttt{android.util.log()} calls in their source code. The percentages further increased to 28.7\% and 40\% for the applications to have more than 50 \texttt{android.util.log()} calls, in benign and malicious applications, respectively. There is no specific indication whether malware samples or benign applications tend to have more such calls than the other. Note that the average number of calls we inserted to craft adversarial examples is only 17 for family mode (refer to Fig. \ref{fig_result_family_jsma}), and 257 for package mode (refer to Fig. \ref{fig_result_package_jsma}). Therefore, the injected code will not bring strong indication that suggests an application to be benign or malicious.

\subsection{Defending Methods}
\label{SubS.DefenceMethods}

\subsubsection{Adversarial training method}
\begin{figure}
  \centering
  \includegraphics[width=1\columnwidth]{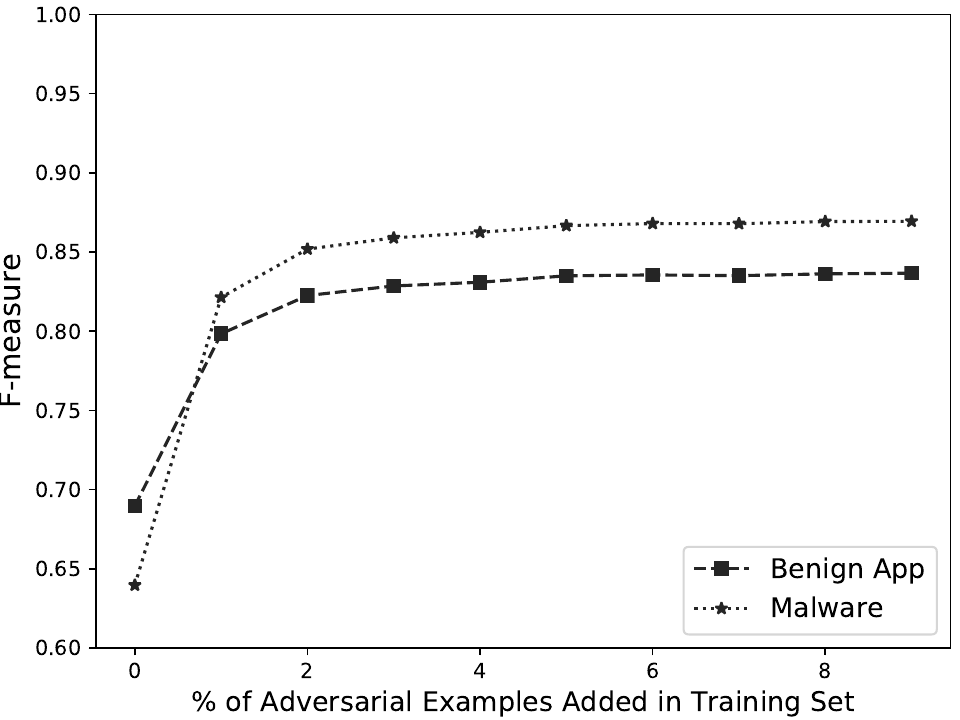}
  \caption{F-measure of benign and malicious applications with the proposed adversarial training defending method on MaMaDroid, with varying percentage of adversarial malware samples added in the training set. 
}
  \label{fig_def_advtraining} 
\end{figure}

The idea of adversarial training is to recursively feed crafted adversarial examples into the training dataset to strengthen the robustness of the machine learning model. We evaluate the effectiveness of adversarial training method on the proposed attack. Fig. \ref{fig_def_advtraining} shows the F-measure of benign and malicious applications with the proposed adversarial training method, with varying percentage of adversarial malware samples added in the training set. Note that the malware test set contains the equal number of original and adversarial malware samples. The F-measure of benign and malicious samples increased from 69\% to 80\% and from 64\% to 82.5\% by adding 1\% of adversarial examples into the training dataset, respectively. Their F-measure further increased to 83\% and 87\%, respectively, when adding more adversarial examples into the training dataset. Although this defending method is simple and effective, it strongly relies on the a priori knowledge of the attack algorithm. 

\subsubsection{Ensemble learning method}
Ensemble of classifiers is one of the effective defences for black-box adversarial example. Instead of training one classifier using the full feature set and all training samples, a number of sub-classifiers are trained with either a subset of features, or a subset of training samples. The final classification result is then made based on the decision of different sub-classifiers with a specific rule, such as majority vote. 
We demonstrate the effectiveness of ensemble learning defence method against our attack. Two scenarios are considered: scenario F and scenario FTB. In scenario F, the attacker cannot query the target model, therefore he/she is not aware of the existence of the defence mechanism. In scenario FTB, the attacker can query the target model with the defence mechanism implemented, and get its prediction result.  

\begin{figure}
  \centering
  \includegraphics[width=1\columnwidth]{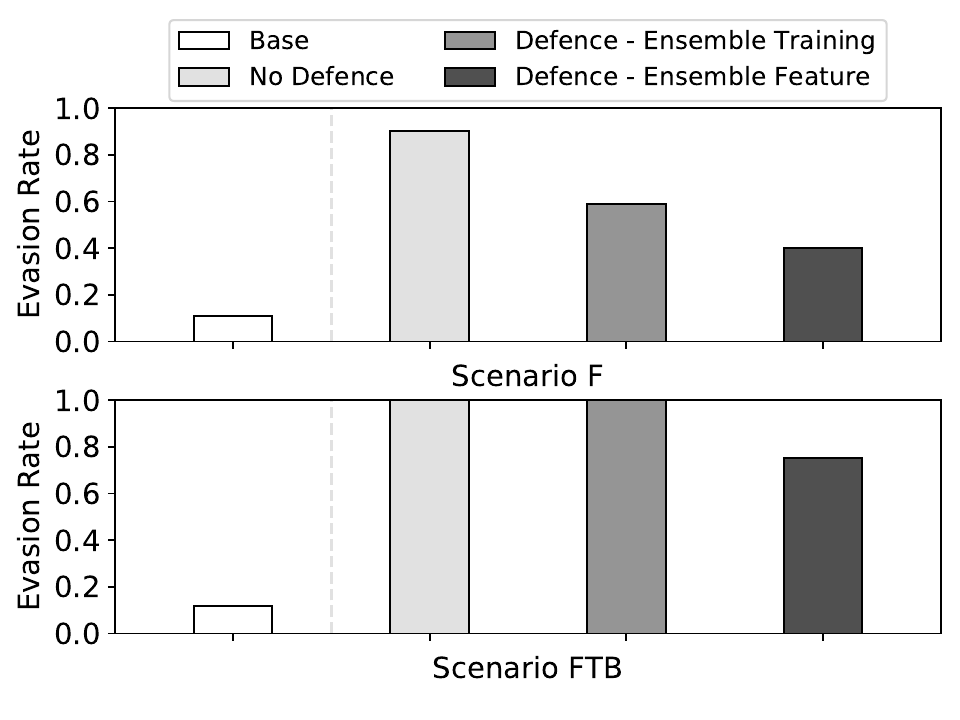}
  \caption{Evasion rate of applying ensemble learning defence mechanism on MaMaDroid family mode in two scenarios (FTB and F). \textbf{Base}: the evasion rate before attack as a baseline; \textbf{No Defence}: attack the model without implementing the defence; \textbf{Defence - Ensemble Training}: attack the model with implementing ensemble learning method as defence, in which each of 10 classifier is trained with 1/10 training samples; \textbf{Defence - Ensemble Feature}: attack the model with implementing ensemble learning method as defence, in which each of 10 classifier is trained with 1/10 features.
}
  \label{fig_ensemble} 
\end{figure}

Fig. \ref{fig_ensemble} presents the results of applying ensemble learning method to defend our C\&W attack on MaMaDroid (family mode), in aforementioned two scenarios, respectively. Two ensemble strategies are implemented in both scenarios, in which each of 10 classifiers is trained with either 1/10 training samples, or 1/10 of features. Evasion rates with and without defence are reported. The results suggest that the ensemble learning method is effective in defending the attack when the attacker has least knowledge of the target model (i.e., scenario F), where the evasion rate decreased from 90\% to 40-59\% in different ensemble strategies. However, when the attacker is capable to query the target model (e.g., the model is implemented as an online service), the defence method cannot effectively defend the proposed attack.

\section{Related Works}
\label{related_work}
\subsection{Adversarial Attacks to Malware Detection}

Recently, there are some research works that studied the security aspect of various machine learning based malware detectors. We give them a brief overview as follows:

Srndic et al. \cite{laskov2014practical} proposed an attack  against PDFrate, an online malicious PDF file detection system. They modified the fields in PDF file that was not rendered by PDF readers. They are extracted as features to discriminate malicious files from benign ones. Similar work was done by Biggio et al. \cite{biggio2013evasion}, who leveraged gradient descent attack to evade detection. Due to the relative simplicity of the PDF file structure, it is easy to alter the file without changing the original content.

Rosenberg et al. \cite{rosenberg2017generic} proposed a black-box attack against machine learning based malware detectors in Windows OS based on analysing API calls. The attack algorithm iteratively added no-op system calls (which are extracted from benign softwares) to the binary code. The proposed method could only be applied to the detection systems that embedded the call sequence into a feature vector. It could not work if the features are statistical information extracted from the call sequence, such as similarity score or probability.  

Grosse et al. \cite{grosse2017adversarial} extended an existing adversarial example crafting algorithm to the Android domain. They trained a deep feed-forward neural network classifier with the feature set adopted in Drebin. It had a comparable detection performance with Drebin. Then, they  launched a white-box attack on the DNN model. In our work, we further customised the algorithm they proposed, and demonstrated a successful black-box attack on the original Drebin system.

Chen et al. \cite{chen2018automated} proposed a poisoning attack for Android malware detection systems through polluting the training set of the original detectors. However, to inject tainted samples into the training set is an arguable assumption in real world scenarios. 
Hu et al. \cite{hu2017generating} demonstrated a generative adversarial network-based (GAN) approach to craft adversarial examples of malware.

In additions, the works \cite{grosse2017adversarial,chen2018automated,hu2017generating} used binary features to indicate the presence of a certain permission or API. The modification on these features usually cannot affect the functionality of the applications. For instance, the adversary can request a new permission in the manifest but will not implement it in the code. Most of recent works will adopt semantic features such as the ones extracted from the control flow graphs. They usually require more cautions to tamper with if we want the application functionality not to be affected. 

\subsection{Android Malware Detection}
Researchers have developed many Android malware detection methods in the last decade. So far, there are a few survey published in this field. Readers could refer to these surveys for typical methods \cite{faruki2015android,suarez2014evolution,la2013survey}. In this subsection, we mainly focus on those which were published recently and used machine learning techniques as their core algorithms.

In this field, almost all recently proposed detectors relied on semantic features to model malware behaviours. For example, Fan et al. \cite{fan2017dapasa} proposed DAPASA, an approach to detect Android piggybacked applications through sensitive subgraph analysis. Xu et al. \cite{xu2016iccdetector} leveraged the inter-component communication patterns to detect Android malware. Yang et al. \cite{yang2014droidminer} developed DroidMiner to scan suspicious applications to determine when they contain malicious modalities. DroidMiner can also be used to diagnose the malware family. Similar idea has also been developed by Li et al. in the work \cite{li2015detection}. Du et al. adopted community structures of weighted function call graphs to detect Android malware. Zhang et al. \cite{zhang2014semantics} proposed a semantic-based approach to classify Android malware via dependency graphs. Gascon et al. \cite{gascon2013structural} developed a method based on efficient embeddings of function call graphs with an explicit feature map. Furthermore, Yang et al. \cite{SY2015} considered user-event-driven components and the related sequences of callbacks from the Android framework to the application code. They further developed a program representation to capture those callback sequences so as to differentiate Android malware from benign applications. 

As explained in Section \ref{intro}, existing works \cite{CL2017,grosse2017adversarial,chen2018automated,demontis2017yes,YW2017} will not work properly when recent detectors relied more on semantic features. In this paper, we presented an advanced method of crafting adversarial examples by applying perturbations directly on the APK \texttt{classes.dex} file. The generated adversarial examples will also be effective on recent detectors that rely more on semantic features.

\section{Conclusion and Future Work}
\label{conclusion}
Recent studies in adversarial machine learning and computer security have shown that, due to its weakness in battling against adversarial examples, machine learning could be a potential weak point of a security system \cite{BM2006,BM2010,WW2018}. This vulnerability may further result in the compromise of the overall security system. 
The underlying reason is that machine learning techniques are not originally designed to cope with intelligent and adaptive adversaries, who can manipulate input data to mislead the learning system.

The goal of this work has been, more specifically, to show that adversarial examples can be very effective to Android malware detectors. To this end, we first introduced a DNN based substitute model to calculate optimal perturbations that also comply with the APK feature interdependence. We next developed an automated tool to implement the perturbations onto the source files (\textit{e.g.}, smali code) of a targeted malware sample. According to the evaluation results, the Android malware detection rates decreased from $96\%$ to $0\%$ in MaMaDroid (\textit{i.e.}, a typical detector that uses semantic features). We also tested Drebin (\textit{i.e.}, a typical detector that uses syntactic features but also collects some features from classes.dex). We found Drebin's detection rates decreased from $97\%$ to $0\%$. To the best of our knowledge, our work is the first one to overcome the challenge of targeting recent Android malware detectors, which mainly collect semantic features from APK's `\texttt{classes.dex}' rather than syntactic features from `\texttt{AndroidManifest.xml}'.

Our future work will focus on two areas: defence mechanisms against such attacks and attack modifications to cope with such mechanisms. For this paper, we only present in Section \ref{SubS.DefenceMethods} a brief discussion about the feasibility and effectiveness of an adversarial training and an ensemble learning defending method. In the next stage, we plan to continue the in depth analysis of various defence mechanisms. 
We will also compare between the effectiveness of different substitute models' architectures.


%
\bibliographystyle{abbrv}
\bibliography{androidhiv}





\ifCLASSOPTIONcaptionsoff
  \newpage
\fi

\end{document}